\documentclass[12pt]{article}
\usepackage{amsfonts, latexsym, amssymb}
\newenvironment{proof}{\textbf{Proof.}~~}{$\Box$}
\newcommand{\im}[0]{\mathrm{im}}

\newcommand{\prfact}[0]{\mathtt{prFact}}
\newcommand{\ssts}[0]{\Sigma^* \times \Sigma^* \rightarrow \Sigma^*}
\newcommand{\stss}[0]{\Sigma^* \rightarrow \Sigma^* \times \Sigma^*}
\newcommand{\sts}[0]{\Sigma^* \rightarrow \Sigma^*}
\newcommand{\ntn}[0]{\mathbb{N} \rightarrow \mathbb{N}}
\newcommand{\getfactors}[0]{\mathtt{getFactors}}
\newcommand{\getproduct}[0]{\mathtt{getProduct}}

\newcommand{\myif}[0]{\mathrm{if}\;}

\def\wit#1{{\mbox{\rm{}WIT}_M(#1)}}
\newcommand{\algorithm}[1]{\parbox{12cm}{\rule[-5mm]{10cm}{.3mm} #1\rule[5mm]{10cm}{.2mm}}}
\makeatletter
\newcommand{\nicesixspacing}{\let\CS=\@currsize\renewcommand{\baselinestretch}{1.6}\tiny\CS}
\newcommand{\nicetenspacing}{\let\CS=\@currsize\renewcommand{\baselinestretch}{2.0}\CS}
\makeatother 
\begin{document}
\sloppy
\begin{titlepage}
\title{Low Ambiguity in Strong, Total, Associative, One-Way Functions}
\author{Christopher M. Homan\thanks{Email: \textbf{choman@cs.rochester.edu}.  Supported in part by grants NSF/CRCD-EEC-98-13002 and NSF-INT-9815095/\protect\linebreak[0]DAAD-315-PPP-g{\"u}-ab.}\\
	Department of Computer Science\\
	University of Rochester\\
	Rochester, NY 14627\\
\\
Univ. of Roch. Dept. of Computer Science, Technical Report 734}
\maketitle
\end{titlepage}
\newtheorem{definition}{Definition}[section]
\newtheorem{theorem}[definition]{Theorem}
\newtheorem{scratch}[definition]{Scratch Definition}
\newtheorem{corollary}[definition]{Corollary}
\newtheorem{lemma}[definition]{Lemma}
\newtheorem{sideeffect}[definition]{Side Effect}
\newtheorem{proposition}[definition]{Proposition}
\abstract{Rabi and Sherman  \cite{rab-she:j:aowf} present a cryptographic paradigm based on associative, one-way functions that are strong (i.e., hard to invert even if one of their arguments is given) and total.  Hemaspaandra and Rothe \cite{hem-rot:j:aowf} proved that such powerful one-way functions exist exactly if (standard) one-way functions exist, thus showing that the associative one-way function approach is as plausible as previous approaches.  In the present paper, we study the degree of ambiguity of one-way functions.  Rabi and Sherman showed that no associative one-way function (over a universe having at least two elements) can be unambiguous (i.e., one-to-one).  Nonetheless, we prove that if standard, unambiguous, one-way functions exist, then there exist strong, total, associative, one-way functions that are $\mathcal{O}(n)$-to-one.  This puts a reasonable upper bound on the ambiguity.  Our other main results are:
\begin{enumerate}
\item P $\neq$ FewP if and only if there exists an $(n^{\mathcal{O}(1)})$-to-one, strong, total AOWF.
\item No $\mathcal{O}(1)$-to-one total, associative functions exist in $\ssts$.
\item For every nondecreasing, unbounded, total, recursive function $g : \ntn$, there is a $g(n)$-to-one, total, commutative, associative, recursive function in $\ssts$.
\end{enumerate}

}
\section{Introduction}\label{intro}

Rabi and Sherman \cite{rab-she:j:aowf} describe protocols for two-party secret-key agreement (due to Rivest and Sherman) and for digital signatures that use strong (i.e., 2-ary, one-way functions that are hard to invert, even if one of their arguments is given), total, associative, one-way functions as cryptographic primitives.  Hemaspaandra and Rothe \cite{hem-rot:j:aowf} prove that such powerful one-way functions exist exactly if (standard) one-way functions exist, thus showing that the associative one-way function approach is as plausible as previous approaches.

In this paper, we study the ambiguity of one-way functions.  Rabi and Sherman showed that no total, associative, one-way function (over a universe having at least two elements) can be unambiguous (i.e. one-to-one).  We strengthen this result in our domain of interest by proving that no total, associative function in $\ssts$ is $\mathcal{O}(1)$-to-one.  Nonetheless, we prove that, if standard (i.e., 1-ary), unambiguous, one-way functions exist, then there exist strong, total, associative, one-way functions that are $\mathcal{O}(n)$-to-one, thereby putting a reasonable upper bound on the ambiguity. 

This paper is organized as follows: in Section 3, we prove---as mentioned above---that no total, associative, function in $\ssts$ is $\mathcal{O}(1)$-to-one.  In addition, we prove that, for every nondecreasing, unbounded, total, recursive function $g$ there exists a $g(n)$-to-one total, associative, \textit{commutative} recursive function in $\ssts$.  In Section 4, we prove that, if standard, unambiguous, one-way functions exist, then $\mathcal{O}(n)$-to-one, strong, total, associative, one-way functions exist, and that FewP $\neq$ P exactly if $n^{\mathcal{O}(1)}$-to-one, strong, total, associative one-way functions exist.  In Section 5, we prove a lower bound on the ambiguity of the class of total, associative functions in $\ssts$ whose output strings are polynomially bounded with respect to their inputs (note that strong, total, associative, one-way functions are a subclass of this class).  Finally, Section 6 presents the conclusion and poses open questions.

\section{Preliminaries}\label{prelim}
Fix the alphabet $\Sigma = \{0,1\}$, and let $\Sigma^*$ denote the set of all strings over $\Sigma$.  We denote the set of all real numbers by $\mathbb{R}$ and the set of all natural numbers (i.e., integers greater than or equal to zero) by $\mathbb{N}$.  

For any two sets $S$ and $T$, $S \times T$ is the set $\left\{(s,t) \;\mid\; (s \in S) \wedge (t \in T)\right\}$.  We use $\prod_{i=1}^n S_i$ as shorthand for $S_1 \times \cdots \times S_n$.

We define $\cup$ over both subsets and multisets of $\Sigma^*$ (a multiset is a set in which multiple instances of the same element may appear).  If $A$ and $B$ are both sets, then $A \cup B$ is the union of $A$ and $B$.  If $A_M$ and $B_M$ are multisets, then $A_M \cup B_M$ is the multiset that contains exactly all of the instances of all the elements of $A_M$ and $B_M$ and nothing else.  If $A$ is a (multi)set, $\|A\|$ is the cardinality of $A$.  For all sets $A$, we define $\mathcal{M}(A)$ to be the set of all multisets whose elements are members of $A$ (a.k.a the ``power multiset'' of $A$).  We will sometimes write a set as $\{a_1, \ldots, a_n\}$ where $a_1, \ldots, a_n$ are its elements, and we will write a multiset as $\{a_1, \ldots, a_n\}_M$, where $a_1, \ldots, a_n$ are its (possibly not distinct) elements.  We may encode a set or multiset as a single string, using some recursive, recursively-invertible, one-to-one function.  For example, we can order the elements of the (multi)set, double each character of each element (except for $\epsilon$, which we denote as $10$), and separate each element with $01$.  

Throughout this paper, we will use ``$\log x$'' to mean ``$\log_2 x$.''

A language $L \subseteq \Sigma^*$ is in UP \cite{val:j:checking} if and only if there exists a nondeterministic Turing machine $M$ that accepts $L$, runs in polynomial time, and has for all inputs at most one accepting path.  A language $L \subseteq \Sigma^*$ is in FewP \cite{all-rub:j:print} if and only if there exists a polynomial $p$ and a nondeterministic Turing machine $M$ that accepts $L$, runs in polynomial time, and on each input $s \in \Sigma^*$ has at most $p(|s|)$ accepting paths.

Let $f:A \rightarrow B$ denote the function $f$, where $A$ is the $\textit{domain}$ of $f$ and $B$ is the $\textit{range}$ of $f$.  A function is $\textit{total}$ if it is defined on each element in its domain.  The $\textit{image}$ of $f$, denoted as $\im(f)$, is the set $\{b \in B \;\mid\; (\exists a \in A)[f(a) \mbox{ is defined and equal to } b]\}$.  The \textit{preimage set} of $b \in B$, denoted $f^{-1}(b)$ is $\{a \in A\;\mid\; f(a) \mbox{ is defined and equal to } b\}$.   A function $g : B \rightarrow A$, \textit{inverts} $f$ if and only if, for all $b \in \im(f)$, $g(b)$ is defined, $f(g(b))$ is defined, and $f(g(b)) = b$.  We say that $f:A\rightarrow B$ is FP-\textit{invertible} if and only if there exists a function $g : B \rightarrow A$ such that $g$ inverts $f$ and $g \in \mbox{FP}$.

Throughout this paper, we use the phrase ``2-ary function'' to mean ``two-argument function'' and the phrase ``1-ary function'' to mean ``one-argument function.''  Unless explicitly stated as being partial, all 2-ary functions are total over $\Sigma^* \times \Sigma^*$.  For any 2-ary function $\sigma$, we will interchangeably use prefix and infix notation, i.e., $\sigma(x,y) \equiv x \sigma y$.

We will sometimes encode pairs of strings as a single string, using some standard, total, bijective, polynomial-time computable pairing function $\langle \cdot, \cdot \rangle:\Sigma^* \times \Sigma^* \rightarrow \Sigma^*$ that has polynomial-time computable inverses and is nondecreasing in each argument when the other argument is fixed. 

A function $f:A \rightarrow \Sigma^*$ is \textit{unbounded} if, for all $n \in \mathbb{N}$, there exists an $s \in A$ such that $|f(s)| > n$. 

Grollman and Selman \cite{gro-sel:j:complexity-measures}
 (see also Ko's independent work \cite{ko:j:operators}) provided the first independent study of complexity-theoretic 1-ary, one-to-one one-way functions.  Definition \ref{owf} below is the standard definition of a (complexity-theoretic) one-way function \cite{gro-sel:j:complexity-measures} for the case of $2$-ary functions that are not one-to-one \cite{rab-she:j:aowf}.

 \begin{definition}\label{owf} \cite{rab-she:j:aowf, hem-rot:j:aowf}~~Let $\sigma:\Sigma^* \times \Sigma^* \rightarrow \Sigma^*$ be an arbitrary 2-ary function.
 \begin{enumerate}
 \item We say $\sigma$ is  \textup{honest} if and only if there exists some polynomial $p$ such that for every $z \in \im(\sigma)$ there exists a pair $(x,y) \in \sigma^{-1}(z)$ such that $|x| + |y| \leq p(|z|)$.
 \item We say $\sigma$ is a \textup{one-way function} if and only if $\sigma$ is honest, polynomial-time computable, and not \textup{FP}-invertible.
 \end{enumerate}
 \end{definition}

 \begin{definition}\label{def:associativity} \cite{hem-rot:j:aowf, rab-she:j:aowf}~~Let $\sigma: \Sigma^*\times\Sigma^* \rightarrow \Sigma^*$ be any total 2-ary function.  We say $\sigma$ is \textup{associative} if and only if $x \sigma (y \sigma z) = (x \sigma y) \sigma z$.
 \end{definition}  

 Actually, Rabi and Sherman \cite{rab-she:j:aowf} deal only with a notion known (in the nomenclature of Hemaspaandra and Rothe \cite{hem-rot:j:aowf}) as \textit{weak associativity}, while Hemaspaandra and Rothe deal with both weak associativity and associativity.  Definition \ref{def:associativity} is that of associativity, but the difference between the two notions is not relevant for us since for total functions the two notions are known to coincide \cite{hem-rot:j:aowf}.

 \begin{definition} \cite{hem-rot:j:aowf,rab-she:j:aowf}~~A total 2-ary function $\sigma:\Sigma^* \times \Sigma^* \rightarrow \Sigma^*$ is an \textup{associative one-way function (AOWF)} if and only if $\sigma$ is both associative and one-way.
 \end{definition}

 \begin{definition} \cite{hem-rot:j:aowf, rab-she:j:aowf}~~A total 2-ary function $\sigma$ is said to be \textup{strong} if and only if $\sigma$ is not \textup{FP}-invertible, even if one of its arguments is given.  More formally, a 2-ary function $\sigma$ is \textup{strong} if and only if neither (a) nor (b) holds:
   \begin{enumerate}
   \item[(a)] There exists a function $g_1 \in \mathrm{FP}$ such that for every $z \in \im(\sigma)$ and for each $x \in \Sigma^*$, if $\sigma(x,y) = z$ for some $y \in \Sigma^*$, then $g_1(\langle x, z \rangle)$ is defined and $\sigma(x, g_1(\langle x, z \rangle)) = z$.
   \item[(b)] There exists a function $g_2 \in \mathrm{FP}$ such that for every $z \in \im(\sigma)$ and for each $y \in \Sigma^*$, if $\sigma(x,y) = z$ for some $x \in \Sigma^*$, then $g_2(\langle y, z \rangle)$ is defined and $\sigma(g_2(\langle y,z\rangle), y) = z$.
   \end{enumerate}
 \end{definition}

 It is known that, unless P $\neq$ NP, some strongly noninvertible functions are invertible \cite{hem-pas-rot:inprep:strong}.  We now define bounded ambiguity for functions over strings.
 \begin{definition} Let $h:\mathbb{N} \rightarrow \mathbb{N}$.  We say a function $\sigma: (\prod_{i = 1}^k \Sigma^*) \rightarrow \Sigma^*$ is \textup{$h(n)$-to-one} if and only if
 \[(\forall y \in \im(\sigma))[ \|\{ x \in \prod_{i = 1}^k \Sigma^*\;\mid\;\sigma(x) = y \}\| \leq h(|y|)].\]
 \end{definition}

 %
 %
 \section{Total, Associative Functions}\label{total}
 In this section we significantly raise the known lower bounds on the ambiguity of total, associative functions in $\Sigma^* \times \Sigma^* \rightarrow \Sigma^*$, thereby raising the same bounds for the class of total AOWFs.  Our goal is to prove that no such constant-to-one, total, associative functions exist.  We will first prove a slightly stronger claim, from which our desired result follows immediately.  
 \begin{lemma}\label{nolimit}For every total, associative function $\sigma: \Sigma^* \times \Sigma^* \rightarrow \Sigma^*$ and every $k \in \mathbb{N}$  there exists a string $t \in \Sigma^*$ for which at least one of the following conditions is true
 \begin{enumerate}
 \item[(a)] $\|\{ x \in \Sigma^*\;\mid\; (x \neq t) \wedge (\exists y \in \Sigma^*)[(x,y) \in \sigma^{-1}(t)] \}\| \geq k$.
 \item[(b)] $\|\{ y \in \Sigma^*\;\mid\; (y \neq t) \wedge (\exists x \in \Sigma^*)[(x,y) \in \sigma^{-1}(t)] \}\| \geq k$.
 \end{enumerate}
 \end{lemma}
 \begin{proof}
 We prove the lemma by induction on $k$. Let $\sigma : \Sigma^* \times \Sigma^* \rightarrow \Sigma^*$ be a total, associative function.
 \begin{description}
 \item[\textbf{Basis case (k = 0):}]~~\\For $k=0$, both (a) and (b) above hold trivially.
 \item[\textbf{Basis case (k = 1):}]~~\\Let $x,y \in \Sigma^*$ be such that $x \neq y$.  Since $\sigma$ is total, $(\exists t \in \Sigma^*)[t = x \sigma y]$.  Since $x \neq y$, either $x \neq t$ or $y \neq t$ (or both).  Therefore, for $k = 1$, $\sigma^{-1}(t)$ generates one of the sets that satisfies one of conditions (a) or (b) above.
 \item[\textbf{Induction step:}]~~\\ Let $k \in \mathbb{N}$ such that $k \geq 1$.  Suppose that no set of size greater than or equal to $k+1$ exists that satisfies one of conditions (a) or (b) above for $\sigma$.   By the induction hypothesis, there exists a $t \in \Sigma^*$ such that $\sigma^{-1}(t)$ generates a set of size $k$ that satisfies one of conditions (a) or (b) above.  In this case, suppose that condition (a) is satisfied (the argument for the former case is analogous to the latter).  By the conditions of (a), there exist strings $x_1, \ldots, x_k, y_1, \ldots, y_k \in \Sigma^*$ (where $x_1,\ldots,x_k$ are distinct, and distinct from $t$) such that 
 \[\{(x_1, y_1), \ldots, (x_k, y_k)\} \subseteq \sigma^{-1}(t).\]
 Choose distinct $s_1, \ldots, s_{k^2 + k + 1} \in \Sigma^*$ satisfying 
 \[\{s_1,\ldots, s_{k^2 + k + 1}\}  \cap \{x_1, \ldots, x_k, t\} = \emptyset.\] 
 Since $\sigma$ is associative, for each $i \in \{1, 2, \ldots, k^2 + k + 1\}$,
 \begin{eqnarray}
 (x_1 \sigma y_1) \sigma s_i = \cdots = (x_k \sigma y_k) \sigma s_i & = & x_1 \sigma (y_1 \sigma s_i) = \cdots = x_k \sigma (y_k \sigma s_i)\\
								& = & t \sigma s_i
 \end{eqnarray}
 (the equation on line (2) holds, because, by assumption, for all $j \in \{1, \ldots, k\}$, $x_j \sigma y_j = t$).  Set $u_i = t \sigma s_i$.  If at least one such $u_i$ is not a member of $\{x_1, \ldots, x_k, t\}$, then $\{x_1, \ldots, x_k, t\}$ satisfies case (a) for $u_i$ and thus contradicts our assumption that no such set of size $k+1$ exists.  Otherwise, every such $u_i$ is a member of $\{x_1, \ldots, x_k, t\}$.  Since $k^2 + k + 1 = (k+1)k + 1$, by the pigeonhole principle, there exists some $t' \in  \{x_1, \ldots, x_k, t\}$ such that
 \[ \|\{ j \in \{1, 2, \ldots, k^2 + k + 1\} \;\mid\; u_j = t'\} \| \geq k+1.\]
 Let $A = \{j \in \{1, \ldots, k^2 + k + 1\} \;\mid\; u_j = t'\}$, and observe that $\|A\| \geq k+1$ and for each $a \in A$,
 \[s_a \in \{y \in \Sigma^*\;\mid\; (y \neq t') \wedge (\exists x \in \Sigma^*)[(x,y) \in \sigma^{-1}(t')]\},\]
 Since we chose distinct $s_i$ this set is large enough to contradict our assumption that no such set of size $k+1$ exists.
 \end{description}
 \end{proof}\\
  The theorem below follows immediately.
 \begin{theorem}\label{noconstant}No total, associative, $\mathcal{O}(1)$-to-one function $\Sigma^* \times \Sigma^* \rightarrow \Sigma^*$ exists.\end{theorem}
An interesting side effect of the proof of Lemma \ref{nolimit} is that, in order to create an image element $s$ with preimage size greater than or equal to $k$, we need compose the total, associative function $\sigma$ with itself no more than $k$ times, assuming that we carefully pick the domain elements; in other words, $s$ is the product of no more than $k+1$ ``factors.''   

 \begin{sideeffect}For any total associative function $\sigma:\Sigma^* \times \Sigma^* \rightarrow \Sigma^*$, and for all $k \in \mathbb{N}$ such that $k > 0$, there exists $k' \leq k+1$ and $s_1, \ldots s_{k'} \in \Sigma^*$ such that $\|\sigma^{-1}(s_1 \sigma \cdots \sigma s_{k'})\| \geq k$.
 \end{sideeffect}
 We will use this result in Section \ref{FIPSPACE}, where we provide a lower bound on the ambiguity of all total, associative functions in $\ssts$ whose output string lengths are polynomially bounded by the length of their corresponding input strings.

 We now prove that, for every nondecreasing, unbounded, total, recursive function $g : \ntn$, there is a $g(n)$-to-one, total, commutative, associative, recursive function in $\ssts$.  
 \begin{theorem}\label{tight}For every nondecreasing, unbounded, total, recursive function $g: \ntn$, there is a $g(n)$-to-one, total, commutative, associative, recursive function $\sigma:\Sigma^* \times \Sigma^* \rightarrow \Sigma^*$, thus placing an upper bound on the ambiguity of this class of functions.
 \end{theorem}
 \begin{proof}
 Let $g : \mathbb{N} \rightarrow \mathbb{N}$ be a nondecreasing, unbounded, total, recursive function.  We will construct a $g(n)$-to-1, total, commutative, associative, recursive function $\sigma:\Sigma^* \times \Sigma^* \rightarrow \Sigma^*$.  Our construction uses a downward self-reducible trick that results in a total, single-valued, one-to-one function $\prfact:\Sigma^* \rightarrow \mathcal{M}(\Sigma^* \setminus \im(\sigma))$ (recall that $\mathcal{M}(\cdot)$ is the ``power multiset'' of $\cdot$) with the following property:
 \begin{quote}
 $s \in \im(\sigma)$ if and only if $s_1 \sigma \cdots \sigma s_k = s$, where $\{s_1, \ldots s_k\}_M = \prfact(s)$.
 \end{quote}
 Since $\sigma$ is associative and commutative, all elements in $\sigma^{-1}(s)$ are of the form $(s_{\pi(1)} \sigma \cdots \sigma s_{\pi(i)}) \sigma (s_{\pi(i + 1)} \sigma \cdots \sigma s_{\pi(k)})$, where $\pi$ is a permutation of $\{1, \ldots, k\}$.  It follows from simple combinatorics that $\|\sigma^{-1}(s)\| \leq \sum_{i = 1}^{k-1}{k \choose i} = 2^k - 2$.  Conversely, $x \sigma y = s$ if and only if $\prfact(x) \cup \prfact(y) = \{s_1, \ldots s_k\}_M$ ($\prfact$ is so named because the properties mentioned above are very similar to certain properties that prime factorizations have over the natural numbers).  Thus, if $\sigma$ can first compute $\prfact(x)$ and $\prfact(y)$ before it computes $s$, it can choose a value for $s$ so that $s$ satisfies the ambiguity bound $g$.

 This can be done as follows:  on input $(a,b)$, $\sigma$ performs the following two phase process.  The first phase starts with an empty set $K$ (so named because it contains the portion of $\prfact$ that is currently ``known''), to which $\sigma$ will add as elements ordered pairs in a well-defined order that is independent of the values $(a,b)$.  In effect, $K$ at any time $t$ constitutes a partial definition of $\prfact$.  We will denote partial function defined by $K$ for time $t$ of $\sigma$ running on input $(a,b)$ as $\prfact_{t,a,b}$, i.e.,
 \begin{eqnarray}
   \prfact_{t,a,b}(x) = \left\{\begin{array}{ll} X_M, & \mbox{ if $\langle x, X_M\rangle \in K$ at time step $t$ of $\sigma$ running on input $(a,b)$}\\
	 \mbox{undefined,} & \mbox{otherwise.}\end{array}\right.
 \end{eqnarray}
 Phase one concludes at some time $t$ such that both $\prfact_{t,a,b}(a)$ and $\prfact_{t,a,b}(b)$ are defined.  If, at time step $t$, there exists a $z \in \Sigma^*$ such that $\prfact_{t,a,b}(z)$ is defined and equal to $\prfact(a) \cup \prfact(b)$, then $\sigma$ outputs $z$.  Otherwise, $\sigma$ chooses $x \in \Sigma^*$ so that 
 \begin{enumerate}
 \item $\prfact_{t,a,b}(x)$ is not defined, and
 \item $g(|x|) > 2^{\|\prfact_{t, a, b}(a) \cup \prfact_{t, a, b}(b)\|} - 2$.
 \end{enumerate}
 $\sigma$ then adds $\langle x,\prfact_{t, a, b}(a) \cup \prfact_{t, a, b}(b)\rangle$ to $K$, outputs $x$, and halts.  

 The partial functions $\prfact_{t, a ,b}$ are, in a sense, analogous to the the stages of a finite extension construction used in relativization proofs (in fact our construction is in some sense a diagonalization of the ambiguity bound $g$---one that is computable, of course).  In order for these partial functions to ``add up'' to a single function (i.e. $\prfact$) that has all the properties we desire, it is crucial that for each pair of input strings $\langle a, b \rangle$ and time step $t$, the definition of $\prfact_{t,a,b}$ is \textit{consistent} with all other $\prfact_{t', a', b'}$ in a significant way.  By this we mean that 
\begin{eqnarray}
(\forall x \Sigma^*)(\exists y \in \Sigma^*)(\forall \prfact_{t, a ,b})[(\prfact_{t, a ,b}(x) \mbox{ is undefined}) \vee (\prfact_{t, a ,b}(x) = y)]
\end{eqnarray}
It is also necessary that every $\prfact_{t,a,b}$ be one-to-one.  We claim that $\sigma$, defined on input $(a,b)$ by the following procedure, gives rise to such a family of functions.

 \algorithm{
 \begin{enumerate}
 \item (Phase one) \textbf{IF} $\langle a, b \rangle \neq \langle \epsilon, \epsilon \rangle$, \textbf{LET} $c = \sigma(a', b')$ (where $\langle a', b' \rangle$ is the string that immediately precedes $\langle a,b \rangle$ in the lexicographical order), and discard $c$.
 \item \textbf{LET} $A_M = \mathtt{getFactors}(a)$,
 \item \textbf{LET} $B_M = \mathtt{getFactors}(b)$,
 \item (Phase two) \textbf{OUTPUT} $\mathtt{getProduct}(A_M \cup B_M)$,
 \end{enumerate}}\\
 where $\mathtt{getFactors}:\Sigma^* \rightarrow \mathcal{M}(\Sigma^*)$, on input $s$, is defined by the following procedure:

 \algorithm{
 \begin{enumerate}
 \item \textbf{IF}, for some $S_M \in \mathcal{M}(\Sigma^*)$, $\langle s, S_M \rangle \in K$, \textbf{OUTPUT} $S_M$,
 \item \textbf{ELSE} \textbf{LET} $K = K \cup \{\langle s, \{s\}_M\rangle\}$, and \textbf{OUTPUT} $\{s\}_M$,
 \end{enumerate}
 }\\
 and on input $A_M \in \mathcal{M}(\Sigma^*)$, $\mathtt{getProduct}: \mathcal{M}(\Sigma^*) \rightarrow \Sigma^*$ is defined by the following procedure:

 \algorithm{
 \begin{enumerate}
 \item \textbf{IF}, for some $z \in \Sigma^*$, $\langle z, A_M \rangle \in K$, \textbf{OUTPUT} $z$, 
 \item \textbf{ELSE} 
 \begin{enumerate}
   \item \textbf{LET} $x = \min\{y \;\mid\; (g(|y|) > 2^{\|A_M\|}-2) \wedge (\forall \langle s, S_M \rangle \in K)[s \neq y]\}$ (where $\min$ is defined relative to the lexicographic ordering),
   \item \textbf{LET} $K = K \cup \{\langle x, A_M\rangle\}$,
   \item \textbf{OUTPUT} $x$.
 \end{enumerate}
 \end{enumerate}
 }\\
 Note that $\getfactors$ and $\getproduct$ are the only places where elements are added to $K$.  Before we prove our claims, we need the following definition:  for all (possibly partial) functions $\alpha$ and $\beta$ defined over the same domain and range, we say that $\alpha$ \textit{extends} $\beta$ if, wherever $\beta$ is defined, $\alpha$ is also defined, and for all $x \in \Sigma^*$ where $\beta(x)$ and $\alpha(x)$ are both defined, $\beta(x) = \alpha(x)$.  Now, from the definition of $\sigma$, the following claims follow easily:
 \begin{enumerate}
 \item For all inputs $a,b \in \Sigma^*$ and at every time step $t$ during the execution of $\sigma$, $\prfact_{t, a, b}$ is one-to-one and single-valued.  This can easily be proved by induction over the lexicographic order of all paired input strings $\langle a, b \rangle$.
 \item For every two pairs of input strings $\langle a, b \rangle$, $\langle a', b'\rangle$, and corresponding time steps $t$ and $t'$, either $\prfact_{t, a,b}$ extends $\prfact_{t', a',b'}$ or $\prfact_{t', a',b'}$ extends $\prfact_{t, a,b}$ (this captures our intuition that the partial functions must be significantly consistent).  This is because the order in which the functions $\getfactors$ and $\getproduct$ are called on particular input values is independent of the input values to $\sigma$ (although, of course, the number of calls in this sequence that are \textit{made} is not), because $\sigma$ never removes elements from $K$, and because the actions that $\getfactors$ and $\getproduct$ take depend only on their respective inputs and on the current value of $K$.
 \end{enumerate}
 Clearly, for every $x \in \Sigma^*$, there are infinitely many partial functions $\prfact_{t, a,b}$ such that $\prfact_{t, a,b}(x)$ is defined, thus any function extending all such $\prfact_{t, a,b}$ must be total.  It follows from item two that there is a unique, single-valued function that extends all partial functions $\prfact_{t, a,b}$.  We will define $\prfact$ to be this unique, total, single-valued function.  We make the following claims:
 \begin{description}
 	\item[Claim 1:]~~\boldmath$(\forall a, b \in \Sigma^*)[(\mathtt{prFact}(a) = \mathtt{prFact}(b)) \Leftrightarrow  (a = b)].$\unboldmath~~\\
Otherwise, since each $\prfact_{t, x, y}$ is one-to-one and single-valued, $\prfact$ would not extend any $\prfact_{t, x, y}$ on which both $a$ and $b$ are defined.  
	\item[Claim 2:]~~\boldmath$(\forall a, b \in \Sigma^*)[\mathtt{prFact}(a \sigma b) = \mathtt{prFact}(a) \cup \mathtt{prFact}(b)].$\unboldmath~~\\
	This follows immediately from the definitions of $\sigma$ and $\prfact$.
 \end{description}
 We are now ready to prove our main claims.
 \begin{description}
 \item[\boldmath$\sigma$ is total:\unboldmath]~~\\Clearly, $\sigma$ halts and outputs on every input, therefore it must be total.   
 \item[\boldmath$\sigma$ is associative:\unboldmath]~~\\For all $a,b,c \in \Sigma^*$, and by claim 2, 
 \begin{eqnarray*}
 \mathtt{prFact}((a\sigma b) \sigma c) &=& \mathtt{prFact}(a\sigma b) \cup \mathtt{prFact}(c)\\
 &=&\mathtt{prFact}(a) \cup \mathtt{prFact}(b) \cup \mathtt{prFact}(c)\\
 &=& \mathtt{prFact} (a) \cup \mathtt{prFact}(b \sigma c)\\
 &=& \mathtt{prFact} (a \sigma (b \sigma c)).
 \end{eqnarray*}  
 By claim 1, $(a\sigma b) \sigma c = a \sigma (b \sigma c)$.
 \item[\boldmath$\sigma$ is commutative:\unboldmath]~~\\For all $a,b \in \Sigma^*$, by claim $2$, $\mathtt{prFact}(a\sigma b) = \mathtt{prFact}(a) \cup \mathtt{prFact}(b) = \mathtt{prFact}(b) \cup \mathtt{prFact}(a) = \mathtt{prFact}(b\sigma a)$.  By claim $1$, $a \sigma b = b \sigma a$.
 \item[\boldmath$\sigma$ is $g(n)$-to-one:\unboldmath]~~\\By claims 1 and 2 above, for all $x \in \im(\sigma)$, and all $a,b \in \Sigma^*$, $(a \sigma b = x) \Leftrightarrow (\mathtt{prFact}(a) \cup \mathtt{prFact}(b) = \mathtt{prFact}(x))$.  There are no more than $2^{\|\mathtt{prFact}(x)\|} - 2$ such pairs $(a,b)$.  Since, for all $\prfact_{t, z, y}$ for which $x$ is defined, we have $\langle x,\mathtt{prFact}(a) \cup \mathtt{prFact}(b)\rangle \in K$ and that $\langle x,\mathtt{prFact}(a) \cup \mathtt{prFact}(b)\rangle$ was added to $K$ during a call to $\getproduct$.  Since, by the construction of $\getproduct$, $g(|x|) > 2^{\|\mathtt{prFact}(x)\|} - 2$, we conclude that $\sigma$ must be $g(n)$-to-one. 
 \end{description}
 We conclude that $\sigma$ is a $g(n)$-to-one, total, commutative, associative, recursive function.
 \end{proof}

 \section{Total, Associative, One-Way Functions}\label{AOWF}
 We now consider the relationship between strong, total, associative, one-way functions and two important complexity classes that frequently appear in the literature on one-way functions.  We will prove that, if P $\neq$ UP, then an $\mathcal{O}(n)$-to-one AOWF exists,
 and that P $\neq$ FewP if and only if an $n^{\mathcal{O}(1)}$-to-one AOWF exists.  Both results follow from the lemma below.
 \begin{lemma}\label{taowf}Let $g : \mathbb{N} \rightarrow \mathbb{N}$ be a function and $L$ be a language accepted by a nondeterministic Turing machine that runs in polynomial time, and, on each input $s$, has at most $g(|s|)$ accepting paths.  If there exists a nondecreasing function $f: \mathbb{N} \rightarrow \mathbb{N}$ such that for all $n \in \mathbb{N}$, $f(n) \geq \max(1, g(n))$, and if $L \not\in$ \textup{P}, then there exists an $\mathcal{O}(n(f(n))^2)$-to-one strong, total AOWF.
 \end{lemma}
 \begin{proof}
 Let $g:\mathbb{N} \rightarrow \mathbb{N}$, $f:\mathbb{N} \rightarrow \mathbb{N}$, and $L$ be as assumed above.  Let $M$ be a nondeterministic Turing machine that accepts 
 $L$, runs in polynomial time, and on input $s$ has no more than $g(|s|)$ accepting paths.  We will use $M$ to build an associative, one-way function $\sigma:\Sigma^* \times \Sigma^* \rightarrow \Sigma^*$ that is strong, total, and $\mathcal{O}(n(f(n))^2)$-to-one.  

 First, we introduce some notation.  Let $a \in
 \Sigma^*$, and let $i \in \mathbb{N}$ be such that $i \geq 1$.  Define $a_{(i)}$
 and $a_{(i+)}$ as follows: if $i \leq |a|$, then $a_{(i)}$ is the
 $i$th character (counting from the left) of $a$, and $a_{(i+)}$ is the substring of $a$
 consisting of all characters in $a$ starting from the $i$th.  If $i >
 |a|$, then $a_{(i)} = a_{(i+)} = \epsilon$.

 We define the set of witnesses for $x \in L$ with respect to $M$ by
 \[\wit{x} = \{w \;\mid\; \mbox{$w$ is a witness for ``$x \in L$''}\}.\]
 Since $M(x)$ has at most $f(|x|)$ accepting paths, $0 \leq \|\wit{x}\| \leq f(|x|)$, and $\|\wit{x}\|= 0$ if and only if $x \not\in L$.  We will assume, without loss of generality, that there exists a strictly increasing polynomial $\rho$ that depends only on $M$ such that for each $x \in L$, and for each $w \in \wit{x}$, $|w| = \rho(|x|)$ and $\rho(|x|) > |x|$.  

 To make $\sigma$ easier to understand, we will construct it from
 several subroutines.  The first plays 
 the role of a ``one-way gate.''  We define the subroutine
 $\gamma:\Sigma^*\rightarrow \Sigma^*$ as follows:
 \[\gamma(d) = \left\{\begin{array}{ll}
     1x & \mbox{if }(\exists x \in L)(\exists w \in \wit{x}) [d = \langle x,w\rangle],\\
     0d &  \mbox{otherwise}.\end{array} \right.\] 
 Clearly, $\gamma$ is total, and for all $t \in \im(\gamma)$, $\|\gamma^{-1}(t)\| \leq f(|t| - 1)$.  
 For $c \in \Sigma^*$, $\beta:\Sigma^*\rightarrow \Sigma^*$ is defined as follows:
 \begin{eqnarray*} 
 \beta(c) = \left\{\begin{array}{ll}
 0\gamma(c_{(2+)}) & \myif c_{(1)} = 1,\\
 111        & \myif c = \epsilon,\\
 00c_{(4+)} & \myif c_{(1)} = 0.
 \end{array}\right.
 \end{eqnarray*}
 Clearly, $\beta$ is total.  Suppose that $e \in \im(\beta)$.  Consider the maximum size of $\beta^{-1}(e)$.  First, from the definition of $\beta$, $e_{(1)}e_{(2)} \in \{00, 01, 11\}$.  Consider each case below:
 \begin{description}
 \item[Case 1:]~\\If $e_{(1)}e_{(2)} = 11$, then $\beta^{-1}(e) = \{\epsilon\}$, therefore $\|\beta^{-1}(e)\| = 1$.  
\item[Case 2:]~\\If $e_{(1)}e_{(2)} = 01$, then $e_{(3+)} \in L$ and $\beta^{-1}(e) = 1\gamma^{-1}(e_{(2+)}) = \{1\langle e_{(3+)},w \rangle \;\mid\; w \in \wit{e_{(3+)}}\}$.  It follows that $\|\beta^{-1}(e)\| \leq f(|e| - 2)$.  
\item[Case 3:]~\\If $e_{(1)}e_{(2)} = 00$, then
$\beta^{-1}(e) \subseteq Z$, where $Z = \{0, 00, 01\} \cup \{0xye_{(3+)}\;\mid\; x,y \in \{0, 1\}\} \cup \{1e_{(3+)}\}$, therefore $\|\beta^{-1}(e)\| \leq 8$.  
\end{description}
We define the 2-ary function $\alpha:\Sigma^* \times \Sigma^* \rightarrow \Sigma^*$ as
\[\alpha(a,b) = 
    0(b_{(1)} \cdot a_{(2)})(a_{(1)} \cdot b_{(2)})a_{(3+)}b_{(3+)},\]
where $\cdot$ is scalar multiplication.  Finally, We define the 2-ary function $\sigma:\Sigma^* \times \Sigma^* \rightarrow \Sigma^*$ as
\[\sigma(s,t) = \alpha (\beta(s), \beta(t))\]
Clearly, $\sigma$ is total and honest.  We claim that $\sigma$ is
      $\mathcal{O}(n(f(n))^2)$-to-one, associative, one-way, and strong.
\begin{description}
\item[\boldmath$\sigma$ is associative:\unboldmath]~~\\Let $s,t,u \in
\Sigma^*$ and $s' = \beta(s), t' = \beta(t), u' = \beta(u)$.  First, observe that 
\begin{eqnarray*}
\beta(s \sigma t) &=& \beta(\alpha(\beta(s), \beta(t)))\\
                  &=& \beta(\alpha(s', t'))\\ 
                  &=& \beta(0(t'_{(1)} \cdot s'_{(2)})(s'_{(1)} \cdot t'_{(2)})s'_{(3+)}t'_{(3+)})\\
&=& 00s'_{(3+)}t'_{(3+)}.
\end{eqnarray*}  
Now, using the above equation where necessary,
\begin{eqnarray*}
(s \sigma t) \sigma u &=& \alpha(\beta(\alpha(\beta(s), \beta (t))), \beta (u))\\
                     &=& \alpha(00s'_{(3+)}t'_{(3+)}, u')\\
                     &=& 0(u'_{(1)} \cdot 0)(0 \cdot u'_{(2)})s'_{(3+)}t'_{(3+)}u'_{(3+)}\\
                     &=& 000s'_{(3+)}t'_{(3+)}u'_{(3+)}\\
                     &=& 0(0 \cdot s'_{(2)})(s'_{(1)} \cdot 0)s'_{(3+)}t'_{(3+)}u'_{(3+)}\\
                     &=& \alpha(s', 00t'_{(3+)}u'_{(3+)})\\
                     &=& \alpha(\beta(s), \beta(\alpha(\beta(t), \beta(u))))\\
                     &=& s \sigma (t \sigma u).
\end{eqnarray*}
\item[\boldmath$\sigma$ is $\mathcal{O}(n(f(n))^2)$-to-one:\unboldmath]~~\\
Suppose that $y$ is in the image of $\sigma$.  It follows that $|y| \geq 3$, 
and that there are exactly $|y| - 2$ pairs of string suffixes $(a_{(3+)},b_{(3+)}) \in \Sigma^* \times
\Sigma^*$ such that $y_{(4+)} = a_{(3+)}b_{(3+)}$.  By the construction of $\sigma$, $y_{(1)} = 0$.
The following table lists all of the possible preimage values $(s, t)$ of $y$, given $y_{(2)}$, $y_{(3)}$, $a = \beta(s)$, and $b = \beta(t)$.  
\scriptsize
\[
\begin{array}{|ll|ll|ll|}
\hline
y_{(2)} & y_{(3)} & b_{(1)} \cdot a_{(2)} & a_{(1)} \cdot b_{(2)} &      s                                                    & t \\
\hline
\hline
0       & 0        & 0 \cdot 0       & 0 \cdot 0        & Z & Z\\

0       & 0       & 0 \cdot 0        & 0 \cdot 1        & Z   & \{1\langle b_{(3+)}, w \rangle \;\mid\; w \in \wit{b_{(3+)}}\}\\
0       & 0       & 0 \cdot 1         & 0 \cdot 0       & \{1\langle a_{(3+)}, w \rangle \;\mid\; w \in \wit{a_{(3+)}}\}    & Z\\
0       & 0       &  0 \cdot 1         &0 \cdot 1       & \{1\langle a_{(3+)}, w \rangle \;\mid\; w \in \wit{a_{(3+)}}\}   & \{1\langle b_{(3+)}, w \rangle \;\mid\; w \in \wit{b_{(3+)}}\}\\
0        &0        &1 \cdot 0        &0 \cdot 1         & Z & \{\epsilon\}\\
0        &0        & 0 \cdot 1        &1\cdot 0         &\{\epsilon\}                                           & Z\\
\hline
1        &0        & 1 \cdot 1         &0\cdot 1       & \{1\langle a_{(3+)}, w \rangle \;\mid\; w \in \wit{a_{(3+)}}\}  &\{\epsilon\} \\
\hline
0        &1        & 0 \cdot 1        &1 \cdot 1         &  \{\epsilon\}                                      & \{1\langle b_{(3+)}, w \rangle \;\mid\; w \in \wit{b_{(3+)}}\}\\
\hline
1        &1        & 1 \cdot 1        & 1 \cdot 1       &   \{\epsilon\}                         & \{\epsilon\} \\
\hline
\end{array}
\]
\normalsize
It is easy to see (by counting the number of distinct elements for a given set of $y_{(2)}y_{(3)}$) that for each $a_{(3+)}$ there are at most $f(|a| - 2)+9$ elements $s$
such that $a = \beta(s)$, and likewise for $b_{(3+)}$.  In sum, then, since $f$ is nondecreasing, there are no more
than $(n - 2)(f(n-2) + 9)^2$ preimage elements $(s, t)$ such that $s \sigma t = y$, so $\sigma$ must be $\mathcal{O}(n(f(n))^2)$-to-one.

\item[\boldmath$\sigma$ is one-way:]~~\\Suppose that there is some polynomial-time
computable function $g:\stss$ that inverts $\sigma$.  We could then decide $L$ in polynomial time as follows:
\begin{quote}
Given any input string $s \in \Sigma^*$, to decide if $s \in L$, compute $g(0011s)$ and accept $s$ if and only if $g(0011s)$ is defined and is equal to $(\epsilon, 1\langle s, w\rangle)$, where $w \in \wit{s}$.
\end{quote}
Therefore, we conclude that $\sigma$ must be one-way.

\item[\boldmath$\sigma$ is strong:]~~\\Suppose that there is some polynomial-time
computable
function $g_1:\sts$ such that for
all strings $c \in \im(\sigma)$, and for all $a \in
\Sigma^*$, if $a\sigma b = c$ for some $b\in\Sigma^*$, then $g_1(\langle a, c\rangle)$ is defined and $a \sigma g_1(\langle a, c\rangle) = c$.  We could then decide $L$ in
polynomial time as follows:
\begin{quote}
Given any input string $s \in \Sigma^*$, to decide if $s \in L$,
compute $g_1(\epsilon, 0011s)$ and accept $s$ if and only if $g_1(\epsilon, 0011s)$ is
defined and is equal to  $1\langle s, w\rangle$, where $w \in \wit{s}$.
\end{quote}
By an
analogous argument, if we assume that there is some function
$g_2:\sts$ such that for all
strings $c$ in the image of $\sigma$, and for all $b \in \Sigma^*$,
if $a\sigma b = c$ for some $a\in\Sigma^*$, then $g_2(\langle b,
c\rangle)$ is defined and $g_2(\langle b, c\rangle)\sigma b = c$, then we arrive at the same contradiction.
\end{description}
\normalsize
We conclude that $\sigma$ is a strong, total, $\mathcal{O}(n)$-to-one,
associative, one-way function.
\end{proof}

The following theorems and corollary follow immediately:

\begin{theorem}
If \textup{P} $\neq$ \textup{UP}, then there exists an $\mathcal{O}(n)$-to-one, strong, total AOWF.
\end{theorem}
\begin{proof}
If $L \in$ UP $-$ P, then $L$ is accepted by a nondeterministic Turing machine that runs in polynomial time and has, at most, one 
accepting path.  Taking $f(n) = g(n) = 1$, by Lemma \ref{taowf} there exists an $\mathcal{O}(n)$-to-one AOWF.
\end{proof}

From Grollmann and Selman's proof that 1-ary, unambiguous one-way functions exist if and only if P $\neq$ UP \cite{gro-sel:j:complexity-measures}, the corollary below follows.
\begin{corollary}
If there exists a 1-ary, unambiguous, one-way function, then there exists an $\mathcal{O}(n)$-to-one, strong, total AOWF.
\end{corollary}
\begin{theorem}\label{fewp}
\textup{P} $\neq$ \textup{FewP} if and only if there exists an $n^{\mathcal{O}(1)}$-to-one, strong, total AOWF.
\end{theorem}
\begin{proof}
For the ``only if'' direction, suppose that $L \notin \mbox{\textup{P}}$ is a language accepted by a nondeterministic Turing machine that runs in polynomial time and, on input $s$, has at most $p(|s|)$ accepting paths (where $p$ is a polynomial).  We can easily find another polynomial $q$ that is nondecreasing and greater than or equal to $\max(1, p)$.  By Lemma $\ref{taowf}$, there exists an $\mathcal{O}(n(q(n))^2)$-to-one strong, total AOWF.  

For the ``if'' direction, if there exists an $(n^{\mathcal{O}(1)})$-to-one, strong, total AOWF $\sigma$, then there exists a 1-ary $(n^{\mathcal{O}(1)})$-to-one one-way function (just compose $\sigma$ with the inverse of a standard pairing function).  Allender \cite[Theorem 6]{all:coutdatedExceptForPUNCstuff:complexity-sparse} proves that FewP $\neq$ P if there exists a (1-ary) $(n^{\mathcal{O}(1)})$-to-one one-way function, therefore FewP $\neq$ P.
\end{proof}

We should point out that Rabi and Sherman \cite{rab-she:j:aowf} describe a multi-party secret key agreement protocol, due to Rivest and Sherman, that uses strong, total, \textit{commutative} AOWFs.  Hemaspaandra and Rothe \cite{hem-rot:j:aowf} prove that strong, total, commutative AOWFs exist exactly if P $\neq$ NP.  Assuming that P $\neq$ UP, we conjecture that their construction could easily be modified to yield strong, total, commutative AOWFs that are constant-to-one for all but one element in the image.  On the other hand, under the same conditions as in Lemma \ref{taowf}, and using similar techniques, we constructed a $2^{\mathcal{O}(n)}$-to-one strong, total, commutative AOWF.  Since this result is not much of a gain, and since the proof is rather technical, we omit it here.

\section{Total, Associative Functions with Polynomially Bounded Outputs}\label{FIPSPACE}
The results of the previous section prove that, under certain common complexity-theoretic assumptions, there are low-ambiguity strong, total AOWFs.  But how low can we go?  From Theorem \ref{noconstant} we know that under no conditions do constant-to-one, total, associative functions exist in $\ssts$.  Here we show how to raise this lower bound when we restrict ourselves to the subclass of this class whose members $\sigma$ have the following property:
  \begin{eqnarray}\label{outbound}
 (\exists \mbox{ polynomial } p)(\forall s_1,s_2\in \Sigma^*)[|s_1 \sigma s_2| < p(\max\{|s_1|,|s_2|\})].
  \end{eqnarray}
Obviously, any lower bound on this subclass is also a lower bound on the subclass of all strong, total AOWFs (assuming they exist).  

Our approach here is straightforward.  We will assume, for the purpose of obtaining a contradiction, that a total, associative function in $\ssts$ exists whose ambiguity is less than the proposed lower bound.  We will then construct an image element of the function, using Corollary \ref{preimage size} and Lemma \ref{image length}, whose preimage set is larger than our assumed lower bound allows.  Corollary \ref{preimage size} follows from the lemma below.

\begin{lemma}
Suppose that $\sigma : \ssts$ is a total, associative function.  For every $k \in \mathbb{N}$ such that $k \geq 1$, there exists a $k' \leq k + 1$ and $s_1, \ldots, s_{k'} \in \Sigma^*$ such that 
\begin{enumerate}\label{factor}
\item $ s_1 \sigma \cdots \sigma s_{k'}$, satisfies condition (a) or (b) from Lemma \ref{nolimit} for $k$,
\item $2 \leq \max\{|s_1|, \ldots, |s_{k'}|\} \leq \lceil 2 \log(k + 1)\rceil$,
\end{enumerate}\end{lemma}
\begin{proof}Let $\sigma$ be an associative function in $\ssts$.  We will prove the above lemma by induction over $k$.  First, assume that $k = 1$.  Clearly, $\epsilon \sigma 00$ satisfies the conditions of the lemma.

Next, suppose that $k\geq 1$.  By the induction hypothesis, there exists $s_1, \ldots, s_{k'} \in \Sigma^*$ such that $s_1 \sigma \cdots \sigma s_{k'}$ satisfies one of conditions $(a)$ or $(b)$ from Lemma \ref{nolimit}, that  $2 \leq \max\{|s_1|, \ldots, |s_{k'}|\} \leq \lceil 2\log(k+1)\rceil$, and that $k' \leq k+1$.  Assume, that, for $k+1$, no $s_{1},\ldots,s_{k'}$ exist with the above properties.  Assume, by the induction hypothesis, and without loss of generality, that $s_1 \sigma \cdots \sigma s_{k'}$ satisfies condition (a) from Lemma \ref{nolimit} (the argument in the case that condition (b) is satisfied is analogous).  By assumption and by the induction hypothesis, the cardinality of the set 
\[S = \{x \in \Sigma^*\;\mid\; (x \neq t) \wedge (\exists y \in \Sigma^*)[(x,y) \in \sigma^{-1}(t)]\}\]
is equal to $k$, where $t = s_1 \sigma \cdots \sigma_{k'}$. We choose the set $T \subsetneq \Sigma^*$ subject to the following constraints
\begin{itemize}
\item $S \cap T = \emptyset$,
\item $\|T\| = k^2 + k + 1$,
\item $(\forall s \in \Sigma^*, t \in T)[s \not\in T \Rightarrow ((|t| \leq |s|) \vee s \in S)]$
\end{itemize}
(the third constraint means that the elements of $T$ are the shortest possible strings that will produce the results desired below).  Clearly, such a $T$ exists.
It follows from the proof of Lemma \ref{nolimit} that for at least one $t' \in T$, the string $s_1 \sigma \cdots \sigma s_{k'} \sigma t'$ satisfies condition (a) or (b) of Lemma \ref{nolimit}.
Also, if $t' \in T$, then $t'$ will be one of the shortest $k + 1 + k^2 + k + 1 = (k + 1)^2 + 1$ strings in $\Sigma^*$.  Thus $|t'| \leq \max\{|t| \;\mid\; t \in T\} \leq \lceil \log((k + 1)^2 + 1) \rceil \leq \lceil \log((k + 2)^2) \rceil = \lceil 2\log(k + 2) \rceil$.  But since by the induction hypothesis $\max\{|s_1|, \ldots, |s_{k'}|\} \leq \lceil 2\log(k + 1)\rceil$, $s_1 \sigma \cdots \sigma s_{k'} \sigma t'$ satisfies condition $2$ above.
\end{proof}

The corollary below follows immediately.

\begin{corollary}\label{preimage size}
Suppose that $\sigma : \ssts$ is a total, associative function.  For every $k \in \mathbb{N}$ such that $k \geq 1$, there exists a $k' \leq k+1$ and $s_1, \ldots, s_{k'} \in \Sigma^*$ such that 
\begin{enumerate}
\item $\|\sigma^{-1}(s_1 \sigma \cdots \sigma s_{k'})\| \geq k$, 
\item $2 \leq \max\{|s_1|, \ldots, |s_{k'}|\} \leq \lceil 2 \log(k + 1)\rceil$,
\end{enumerate}
\end{corollary}

Next, we provide an upper bound on the size of the output of any associative function whose outputs are polynomially bounded by its input sizes.

\begin{lemma}\label{image length} Let $\sigma$ be any total, 2-ary function in $\ssts$.  If $\sigma$ satisfies formula (\ref{outbound}),  
then
  \begin{eqnarray}\label{afupper}
  (\exists j \in \mathbb{N}:j > 1)(\forall k \in \mathbb{N} : k > 1)(\forall s_1, \ldots, s_k \in \Sigma^*)[|s_1 \sigma \cdots \sigma s_k| < (\max\{2, |s_1|, \ldots, |s_k|\})^{j^{\lceil\log k\rceil}}].
  \end{eqnarray}
\end{lemma}
\begin{proof}Suppose that $\sigma$ satisfies formula (\ref{outbound}).  We can write formula (\ref{outbound}) equivalently as
\begin{eqnarray}\label{realoutbound}
(\exists m,i \in \mathbb{N} : i > 0)(\forall s_1,s_2\in \Sigma^*)[(\max\{|s_1|,|s_2|\} > m) \Rightarrow (|s_1 \sigma s_2| < (\max\{|s_1|, |s_2|\})^i)],\nonumber
\end{eqnarray}
We will use induction over $k$ to prove that $j = \max\{i + 1, 1 + \lceil\log(\max\{|x \sigma y| : (|x| \leq m) \wedge (|y| \leq m)\})\rceil\}$ satisfies the conditions of the lemma.  Suppose that $k = 2$.  It follows immediately that, for all $s_1, s_2 \in \Sigma^*$, $|s_1 \sigma s_2| < (\max\{2, |s_1|, |s_2|\})^j$.  

Next, suppose that $k = 3$.  By associativity, 
\begin{eqnarray*}
|s_1 \sigma s_2 \sigma s_3| &=& |(s_1 \sigma s_2) \sigma s_3|\\
                            &<& (\max\{(\max\{2, |s_1|, |s_2|\})^j, |s_3|\})^j\\
                            &\leq& (\max\{(\max\{2, |s_1|, |s_2|, |s_3|\})^j, |s_3|\})^j\\                         
                            &=& ((\max\{2, |s_1|, |s_2|, |s_3|\})^j)^j.                         
\end{eqnarray*}
for our choice of $j$.  Now,
\begin{eqnarray*}
  ((\max\{2, |s_1|, |s_2|, |s_3|\})^j)^j &=& (\max\{2, |s_1|, |s_2|, |s_3|\})^{j^2}\\
                                  &=& (\max\{2, |s_1|, |s_2|, |s_3|\})^{j^{\lceil \log 3\rceil}}.
\end{eqnarray*}
Suppose that $k \geq 3$.  Let $k'$ be a natural number satisfying $k \geq k' > 1$.  By the induction hypothesis, 
\begin{eqnarray*}
  (\forall s_1, \ldots, s_{k'}\in \Sigma^*)[|s_1 \sigma \cdots \sigma s_{k'}| < (\max\{2,|s_1|, \ldots, |s_{k'}|\})^{j^{\lceil\log k'\rceil}}].
\end{eqnarray*}
By associativity, 
\begin{eqnarray*}
\lefteqn{|s_1 \sigma \cdots \sigma s_{k + 1}| = |(s_1 \sigma \cdots \sigma s_{\lfloor \frac{k + 1}{2} \rfloor}) \sigma (s_{\lfloor \frac{k + 1}{2} \rfloor + 1} \sigma \cdots \sigma s_{k + 1})|}\\
& < & (\max\{(\max\{2, |s_1|, \ldots, |s_{\lfloor \frac{k + 1}{2} \rfloor}|\})^{j^{\lceil\log(\lfloor \frac{k + 1}{2} \rfloor)\rceil}},(\max\{2, |s_{\lfloor \frac{k + 1}{2} \rfloor + 1}|, \ldots, |s_{k + 1}|\})^{j^{\lceil\log(\lceil \frac{k + 1}{2} \rceil)\rceil}}\})^j \\
& \leq & ((\max\{2, |s_1|, \ldots, |s_{k + 1}|\})^{j^{\lceil \log (k + 1) \rceil - 1}})^j \\
& = & (\max\{2, |s_1|, \ldots, |s_{k + 1}|\})^{j^{\lceil \log (k + 1) \rceil}}
\end{eqnarray*}
(to see why $\lceil\log(\lceil \frac{k + 1}{2} \rceil)\rceil \leq \lceil \log (k + 1) \rceil - 1$, consider that $\lceil\log(\lceil \frac{k + 1}{2} \rceil)\rceil < \log(\frac{k + 1}{2}) + 1 = \log(k + 1) \leq \lceil\log(k + 1)\rceil$).
\end{proof}

Now, we combine the results of Lemma \ref{image length} and Corollary \ref{preimage size} to prove a lower bound on the ``many-to-one''-ness of functions that satisfy formula (\ref{outbound}).

\begin{theorem}\label{badbound}
   For every total, associative function $\sigma : \ssts$ that satisfies formula (\ref{outbound}), there exists an $l \in \mathbb{N}$ where $l > 1$ such that $\sigma$ is not $o(g(n))$-to-one, where $g:\mathbb{N} \rightarrow \mathbb{N}$ inverts $f:\{r \in \mathbb{R} \;\mid\; r \geq 1\} \rightarrow \mathbb{N}$, defined as $f(n) = \lceil2\log n\rceil^{l^{\lceil\log n\rceil}}$.
\end{theorem}
\begin{proof}Suppose that $\sigma : \ssts$ is a total, associative function that satisfies formula (\ref{outbound}).  By Lemma (\ref{image length}), there exists $j \in \mathbb{N}$ where $j > 1$ such that for all $k \in \mathbb{N}$ where $k > 1$, and all $s_1, \ldots, s_k \in \Sigma^*$, $|s_1 \sigma \cdots \sigma s_k| < (\max\{2, |s_1|, \ldots, |s_k|\})^{j^{\lceil\log k\rceil}}$.  We will prove, by contradiction, that $\sigma$ is not $o(g(n))$-to-1, where $g$ inverts $f:\{r \in \mathbb{R} \;\mid\; r \geq 1\} \rightarrow \mathbb{N}$, defined as $f(n) = \lceil2\log n\rceil^{l^{\lceil\log n\rceil}}$. 

Assume that, for all $l > 1$, $\sigma$ is $o(g(n))$-to-one.  Let $l = \lceil j^2 \rceil$.  By assumption,
\begin{eqnarray}\label{falsebound}
  (\forall \delta \in \mathbb{R})(\exists N \in \mathbb{N})(\forall m > N)\left[\frac{\max\{ \|\sigma^{-1}(s)\| \;\mid\; |s| = m\}}{g(m)} < \delta\right]
\end{eqnarray}
Suppose that $\delta = 1$.  Choose $N \in \mathbb{N}$ such that $N$ satisfies equation (\ref{falsebound}).  Let
\begin{eqnarray}\label{nhash}
  n = 8 + \max\left\{\|\sigma^{-1}(s')\| \;\mid\; |s'| \leq N\right\}.
\end{eqnarray}
By Corollary \ref{preimage size}, for some $n' < n$, there exists  $s_1, \ldots, s_{n'} \in \Sigma^*$ such that
\begin{enumerate}
\item $\|\sigma^{-1}(s_1 \sigma \cdots \sigma s_{n'})\| \geq n - 1$, \label{maxpoint}
\item $2 \leq \max\{|s_1|, \ldots, |s_{n'}|\} \leq \lceil 2\log(n)\rceil$,
\end{enumerate}
Let $m = |s_1 \sigma \cdots \sigma s_{n'}|$.  By equation (\ref{nhash}) and item 1 above, $m > N$.  Since $\|\sigma^{-1}(m)\| \geq n-1$, we have $\max\{ \|\sigma^{-1}(s)\| \;\mid\; |s| = m\} \geq n - 1$.  By Lemma \ref{image length} (and because $\max\{|s_1|, \ldots, |s_{n'}|\} \geq 2$),
\begin{eqnarray*}
  m &\leq& (\max\{|s_1|, \ldots, |s_{n'}|\})^{j^{\lceil\log n'\rceil}} \leq (\max\{|s_1|, \ldots, |s_{n'}|\})^{j^{\lceil\log n\rceil}}.
\end{eqnarray*}
By item 2 above, $\max\{|s_1|, \ldots, |s_{n'}|\} \leq \lceil 2\log(n)\rceil$, therefore
\begin{eqnarray*}
 m &\leq& \lceil 2\log(n)\rceil^{j^{\lceil\log(n)\rceil}}.
\end{eqnarray*}
Now,
\begin{eqnarray*}
f(n - 1) &=& \left\lceil 2 \log (n-1)\right\rceil^{(j^2)^{\lceil\log (n-1)\rceil}},
\end{eqnarray*}
which, since $n \geq 8$, $j > 1$,
\begin{eqnarray*}
&>&\lceil 2\log n\rceil^{j^{\lceil\log n\rceil}}\\
&\geq&m.
\end{eqnarray*}
Since $f$ is nondecreasing,
\begin{eqnarray*}
n-1 &>& g(m)\\
 \max\{ \|\sigma^{-1}(s)\| \;\mid\; |s| = m\} &\geq& n - 1 \; > \;  g(m)\\
 \frac{\max\{ \|\sigma^{-1}(s)\| \;\mid\; |s| = m\}}{g(m)} &>&  1,
\end{eqnarray*}
thus, for $l = \lceil j^2 \rceil$ and $\delta = 1$, and for all $N \in \mathbb{N}$, there exists $m > N$ such that $\frac{\max\{ \|\sigma^{-1}(s)\| \;\mid\; |s| = m\}}{g(m)} > \delta$.  But this contradicts our assumption that $\sigma$ is $o(g(n))$-to-1.
\end{proof}

There still remains a very large gap between the known ambiguity of the class of strong, total AWOFs under various existence assumptions, and the lower bound of this property.  We believe that stronger results are possible. 

\section{Conclusion and Open Problems}
We proved that, if unambiguous one-way functions exist, then we can construct strong, total AOWFs with low ambiguity, and that $n^{\mathcal{O}(1)}$-to-one strong total AOWFs exist exactly if P $\neq$ FewP.   Without appeal to ``one-way''-ness, we proved that no total, associative, recursive function in $\ssts$ is $\mathcal{O}(1)$, and that, for every nondecreasing, unbounded, total, recursive function $g : \ntn$, there exists a $g(n)$-to-one total, associative, \textit{commutative} recursive function in $\ssts$.  Finally, we proved that, for every total, associative function $\sigma$ in $\ssts$ whose output strings are polynomially bounded by the lengths of their corresponding input strings, there exists a natural number $l > 1$ such that the ambiguity of $\sigma$ is not $o(g(n))$-to-one, where $g$ inverts $f:\{r \in \mathbb{R} \;\mid\; r \geq 1\} \rightarrow \mathbb{N}$, defined as $f(n) = \lceil2\log n\rceil^{l^{\lceil\log n\rceil}}$.

We mention two open problems.  First, what is the tight lower bound on the ambiguity of the class of strong, total AOWFs?  Second, are there any conditions under which strong, total, commutative, AOWF exist that have reasonable limits on their ambiguity?  

\textbf{Acknowledgments:}  I am grateful to Lane Hemaspaandra for suggesting this topic and for his continual guidance and encouragement, to Alina Beygelzimer, Lane Hemaspaandra, Harald Hempel, J\"org Rothe, and Mayur Thakur for their careful reviews and numerous suggestions.  The link to FewP in Theorem \ref{fewp} was suggested by Alan Selman.  An interesting alternate proof of Theorem \ref{noconstant} was observed by Edith Elkind.  
\bibliographystyle{alpha} \bibliography{jbib}
\end{document}